\documentclass[pra,preprint]{revtex4}
\usepackage{graphics}
\usepackage{amsfonts}
\usepackage{bm}

\begin{document}
\begin{center}
LOW-ENERGY QUADRUPOLE MODES IN DEFORMED CLUSTERS

\vspace{0.5cm}
V.O. Nesterenko$^{1,2}$, W. Kleinig$^{1,3}$,
P.-G. Reinhard$^{4}$, and D.S. Dolci$^{1}$

\vspace{0.5cm}
$^{1}$
BLTP, Joint Institute for Nuclear Research, Dubna, Moscow region,
141980, Russia\\
E-mail: nester@thsun1.jinr.ru\\
$^{2}$Max Planck Institute for Physics of Complex Systems,
01187, Dresden, Germany\\
$^{3}$Technische Univirsitat Dresden, Inst. f\"ur Analysis,
D-01062, Dresden, Germany\\
$^{4}$Institut f\"ur Theoretische Physik,
Universitat Erlangen, D-91058, Erlangen, Germany
\vspace{1.5cm}

ABSTRACT
\end{center}
\vspace{0.5cm}

Properties of low-energy (infra-red) quadrupole modes
(LEQM) of multipolarity $\lambda\mu =$20, 21 and 22
in deformed sodium clusters are studied within the Kohn-Sham
LDA RPA approach.
Possible manifestations of LEQM in stimulated Raman adiabatic
passage (STIRAP) reaction are discussed.
It is shown that, in free light clusters, where the
low-energy spectrum is delute, LEQM can be unambiguously
identified as particular electron-hole pairs. This
gives a chance to reconstruct the mean field
level scheme near the Fermi surface. Moreover,
due to the connection with electric $\lambda\mu =$21 mode,
the scissors mode can be detected.
In heavy (supported) oblate
clusters, LEQM are in general rather involved.
Nevertheless, some interesting $\lambda\mu =$21 and 22
structures determined by specific deformation  effects
can be resolved. The origin of the structures is discussed in
detail.

\newpage
\section{Introduction}

Collective oscillations of valence electrons in metal
clusters manifest themselves in a variety of electric and
magnetic (orbital) plasmons (see \cite{Ne_SN} for review).
Till now only the electric dipole (E1) plasmon was
thoroughly investigated \cite{Hab}. Other plasmons were
not still observed and our knowledge about them is scares.
At the same time, they promise interesting
physics and so devote careful study.

In the present paper, we consider low-energy electric quadrupole
modes in axially deformed clusters.
Quadrupole excitations of valence electrons in such clusters
are collected into bunches
$\Delta {\mathbb N} =0$ and 2,
where ${\mathbb N}$ is the principle shell quantum number
\cite{Ne_PRA,Ne_AP}.
The low-energy bunch $\Delta {\mathbb N} =0$ corresponds to
E2 transitions inside the valence shell. It is clear that this
bunch exists only in clusters with partly occupied valence shell.
Just these clusters exhibit a non-spherical shape. The
high-energy bunch $\Delta {\mathbb N} =2$ corresponds to E2
transition through two shells. This bunch exists in clusters of
any shape and exhausts most of the quadrupole strength. Though
the bunch $\Delta {\mathbb N} =2$ shares the  energy region
with the dominant E1 plasmon, it can be possibly observed by means
of angular-resolved electron energy-loss spectroscopy (AR-EELS)
at electron scattering angles $\sim 6^{\circ}$ \cite{Ger}.

In this paper, we will consider the low-energy quadrupole modes
(LEQM) with $\Delta {\mathbb N} =0$, whose properties,
by our knowledge, were never earlier investigated. We will show
that LEQM demonstrate spectacular structures connected with
cluster deformation.

LEQM are probably too weak
to be observed in AR-EELS but have a chance to be detected
in two-photon processes like Raman scattering (RS) or resonance
fluorescence (RF) where
the competition with E1 modes is completely excluded. These
processes start from excitation of an intermediate E1
electronic state (virtual or real). Then the coupling of the
intermediate and low-energy states results in the population
of the latter, again through E1 transitions (e.g.,
the electron-phonon coupling results in RS population of
ionic phonon states).

As far as we know, there was only one attempt to observe
LEQM by RS in nanoparticles \cite{Duval}. In that experiment
a broad bump was observed at $\sim 0.12$ eV in supported
silver clusters. The electronic
nature of the bump was justified by the inverse cluster size
dependence of the bump energy. Recent data of
this group hint at the deformation dependence of the bump energy
\cite{Duval_privat}. Explanation of these
data requests a strong theoretical support. Our study could be
a first guide in this new field.

Perhaps, a weak coupling of electron states does not favor
their detection by RS. In this connection, the stimulated Raman adiabatic
passage (STIRAP) seems to be more appropriate. This method uses
partially overlapping pulses (from pump and Stokes lasers) to produce
complete population transfer between two quantum states $|1>$ and $|3>$
through the intermediate state $|2>$ \cite{Berg}. The
combination of the pump and Stokes frequencies has to be
resonant with the two-photon Raman transition. In fact,
the Stokes pulse plays the same role as the coupling
between the states $|2>$ and $|3>$ in RS.

LEQM seem to be most interesting in the cases of free
light non-spherical clusters and supported heavy oblate clusters.
In free light clusters, the low-energy spectrum is very dilute
and thus LEQM can be safely resolved. We will show that
LEQM can be unambiguously identified as particular electron-hole
($1eh$) pairs involving the single-particle electron states
near the Fermi level. This allows to obtain, at least partly,
the single-particle spectrum in the clusters.
Since this spectrum is sensitive to many factors (deformation,
temperature, influence of detailed ionic structure, ...), we
thus handle the effective method to study diverse
cluster's properties. The spectrum can be also used as a
robust test for theoretical models. Besides, due to the
connection between LEQM and scissors mode
\cite{LS_ZPD,Ne_sc,Re_M1,Ne_M1_jell}, the latter can be detected as
well.

Heavy clusters in experiments are usually supported and range
shapes from spherical to strongly oblate \cite{Hab}. Size
and shape of supported clusters can be monitored \cite{Tra}.
If heavy clusters are supported on dielectric (and porous) surfaces,
then the interface effect is mainly reduced to shaping the cluster
\cite{Hab}. Such clusters can be simulated, in a first
approximation, by {\it free} clusters at a certain oblate shape.
In the present paper, we pretend only to a rough estimation of
LEQM properties and so this simulation is reasonable for our aims.

As a simplest case, we will consider sodium clusters. However,
LEQM properties described below are of a general character and
so should also take place in other clusters exhibiting quantum
shells, e.g. in clusters of noble metals.

\section{Calculation scheme}

LEQM are described in the linear regime within the
random-phase-approximation (RPA) method
\cite{Ne_PRA,Ne_AP} based on the Kohn-Sham LDA functional
\cite{GL}. The ions are treated in the soft jellium
approximation. High reliability of the method has been checked
in diverse studies of E1 plasmon in spherical
\cite{Ne_EPJD_98} and deformed \cite{Ne_AP,Ne_EPJD_02}
clusters.

We consider axial prolate Na$^+_{15}$ as an example of a
free light cluster. Equilibrium quadrupole and hexadecapole
deformations of Na$^+_{15}$ ($\delta_2 =$0.57 and $\delta_4 =$-0.18)
are determined by minimization of its total
energy \cite{Ne_AP,Ne_EPJD_02}. The neutral Na$_{118}$
at oblate deformations $\delta_2 =$-0.3 and -0.5 is used
as a simulation of heavy supported clusters.

STIRAP is treated as a resonance fluorescence (RF)
process running through  $\lambda\mu =10$ and 11 branches of
the dipole plasmon, as intermediate states (see Fig. 1).
In strictly deformed nuclei, where the branches are well
separated by energy (see Figs. 2 and 3),
independent measurements through every branch are preferable.
The population picture is then simpler and allows easier
interpretation. For example, if STIRAP runs only through
$\lambda\mu =10$ plasmon branch, then the population of
the low-energy quadrupole 22 mode is forbidden
and thus remaining 20 and 21 modes can be easier treated.

STIRAP population of LEQM is approximately calculated
as a coherent sum of independent two-step processes, every one
being a product of dipole photoabsorption and emission:
\begin{eqnarray}
\label{eq:STIRAP}
&\sigma_{2\mu_2 i_2}&
 =
\sum_{i_1}
\sigma^{ab}_{E1\uparrow} (0 \rightarrow 1\mu_1 i_1) \cdot
\sigma^{em}_{E1\downarrow} (1\mu_1 i_1 \rightarrow 2\mu_2 i_2)
\\
& \Rightarrow &
\sum_{i_1}
\omega_{1\mu_1 i_1} |<Q_{1\mu_1 i_1}|E1|0>|^2
(\omega_{1\mu_1 i_1}-\omega_{2\mu_2 i_2})^3
|<Q_{2\mu_2 i_2}|E1|Q^{\dagger}_{1\mu_1 i_1}>|^2 .
\nonumber
\end{eqnarray}
Here $|1\mu_1 i_1>$ and $|2\mu_2 i_2>$ are
RPA states of the dipole plasmon and LEQM, respectively.
Index $i$ runs all the RPA states of the given multipolarity
in the chosen energy interval.
The operator of the dipole transition is electron-hole
in the photoabsorption and electron-electron (hole-hole) in
the photoemission.

Eq. (\ref{eq:STIRAP}) follows from the general expression
for RS cross-section \cite{Greiner} if one neglects the
interference between the neighbouring RPA states.
This is the case for light clusters (see Fig. 2) but not
for heavy ones whose spectrum is rather dense
\cite{Ne_AP,Ne_EPJD_02}. However, the
interference mainly leads to smoothing the response,
which can be taken into account by the reasonable averaging the
results. In the present paper, we weight the {\it low-energy}
responses by the Lorentz function with the averaging parameter
$\Delta =$0.1 eV. This should simulate both the interference and
temperature smoothing. We emphasize that the structures
discussed below are strong enough to be completely smoothed out
and so the main conclusions of the present study have
to be valid in spite of the interference.

Spread of collective
oscillations is known to increase with the excitation energy.
So, for the dipole plasmon which lies much higher than LEQM,
we use the larger averaging with $\Delta =$0.25 eV. Such averaging
was successfully used in our previous RPA calculations
for the photoabsorption \cite{Ne_EPJD_98,Ne_AP,Ne_EPJD_02}.

\section{Results and discussion}

Main results of the calculation are collected in Fig. 4.
The first line of the figure contains photoabsorption
for low-energy quadrupole states $\lambda\mu =$20, 21 and 22.
Photoabsorption is a
useful step in any analysis of electron modes.
The second line of Fig. 4 exhibits
photoabsorption for the scissors magnetic dipole mode.
The next two lines provide the STIRAP populations
(\ref{eq:STIRAP}) for the cases when the reaction runs
separately through $\lambda\mu =10$ and $\lambda\mu =11$
branches of the dipole plasmon (these cases are marked in Fig. 4
as RF E10 and RF E11, respectively).

In the energy intervals given in Fig. 4 (0-1.5 eV for
Na$^+_{15}$ and 0-1.0 eV for Na$_{118}$), we take
into account all the quadrupole RPA states. As for the
dipole intermediate RPA states used in the STIRAP calculations,
we involved all the states in Na$^+_{15}$ and 40-50 of most
collective states in Na$_{118}$, determined by the
appropriate cut-off of the photoabsorption strength.

\subsection{Light free clusters}

The first plot in the left column of Fig. 3 shows that LEQM
photoabsorption in Na$^+_{15}$ is dominated by two
$\lambda\mu =$21 peaks. Following the level scheme for Na$^+_{15}$
given in Fig. 5, these peaks can be associated with electron-hole
pairs [211]-[202] and [211]-[200]. Our analysis shows that
i) collective shifts for these peaks are very small
and ii) just the above $1eh$ pairs provide dominant
(up to $95\%$) contributions to the normalization
conditions of the peaks. This justifies identification of the
peaks as [211]-[202] and [211]-[200] $1eh$ pairs.
The contributions into the optical response of other quadrupole
modes, 20 and 21, given in Fig. 5  is negligible because
of low values of their transition single-particle matrix elements.

Both [211]-[202] and [211]-[200] $1eh$ pairs include the Fermi level
[211] whose energy can be obtained from the ionization potential
data. Then one immediately gets the energies of electron levels
[202] and [200]. So,  STIRAP measurements
together with other data (ionization potential, photoemission data,
etc) allow to obtain the single-particle spectrum
in light deformed clusters. Since this spectrum is sensitive
to many factors (deformation, temperature, influence of detailed
ionic structure, ...), we can use it to study
diverse cluster's properties. Beside, the spectrum can serve
as a robust test for theoretical models.

The next plot for Na$^+_{15}$ exhibits the photoabsorption
cross section for the scissors mode. We see again two peaks
corresponding to [211]-[202] and
[211]-[200] states discussed above. While the lower
state [211]-[202] favors the M1 response, the higher state
[211]-[200] responds mainly to E21 field.
Both photoabsorption plots display a close connection
(coupling) between the scissors M1 and quadrupole E21 modes.
The modes are characterized by one and the same set
of $1eh$ pairs with quantum numbers $\Lambda^{\pi}=1^+$
(where $\Lambda$ is the projection of the orbital moment to
the symmetry axis $z$ and $\pi$ is the space parity of the state).
The coupling of electric and magnetic modes with the same
quantum numbers $\Lambda^{\pi }$ is a general feature of deformed
quantum systems. For example, this feature is well known in atomic
nuclei \cite{Kva_PRC}.

In fact, both scissors M1 and quadrupole E21 modes represent
one and the same intrinsic electron motion in a deformed cluster.
The scissors mode is more specific, while the E21 mode is more
general. To illustrate this point, we expand the single-particle
wave functions in a deformed mean field in terms of the spherical
basis  $(n L\Lambda )$
\begin{equation}
 \Psi_{\nu =[{\cal N}n_z\Lambda ]}
 =
 \sum_{nL} a^{\nu}_{nL}R_{nL}(r) Y_{L\Lambda}(\Omega).
\end{equation}
This allows to evaluate the single-particle orbital M1 transition
amplitude between hole ($\nu =h$) and particle ($\nu =p$) states
\begin{equation}\label{eq:me}
 \langle\Psi_{p}|{\hat L}_{x}|\Psi_{h}\rangle \propto
\delta^{\mbox{}}_{\pi_{p},\pi_{h}}\delta^{\mbox{}}_{\Lambda_{p},
\Lambda_{h}\!\pm\!1}
\sum_{nL}
a^{p}_{nL}a^{h}_{nL}\sqrt{L(L\!+\!1)\!-\!\Lambda_h(\Lambda_h\!\pm \!1)}.
 \nonumber
\end{equation}
Eq. (\ref{eq:me}) shows that the scissors mode is generated by
$\Lambda_p=\Lambda_h\pm 1$ transitions between the components of one
and the same spherical $(nL)$-level. Such selectivity is explained by
the fact that the scissors operators, ${\hat L}_{x}$ and
${\hat L}_{y}$, do not depend
on the space coordinate $r$ and so, due to orthogonality of the radial
wave functions $R_{nL}(r)$, cannot connect the components originating
from different $(nL)$-subshells. Instead, the quadrupole operator
$r^2 Y_{21}$ does connect the components with different
$(nL)$. So, the scissors operators are more selective than the
E21 operator. Every $\Lambda^{\pi}=1^+$ state responds
to both M1 and E21 external fields. Magnitudes of the
responses depend on the wave function of the state.

The third and forth plots show STIRAP populations when the
reaction runs through $\lambda\mu =10$ and $\lambda\mu =11$
branches of the dipole plasmon (RF E10 and
E11 cases, respectively). Like in the photoabsorptions,
$\lambda\mu =21$ mode dominates over 20 and 22.
In RF E10 case, the populations of both
[211]-[202] and [211]-[200] $1eh$ states are equally strong.
Detection of these states provides energies of
electron levels [202] and [200]. Besides,
the scissors mode associated with the state [211]-202]
can be observed. In RF E11 case, only [211]-[200] $1eh$
state is well populated. The differences between
RF E10 and E11 cases can be of use (for example, to
distinguish [211]-[202] and [211]-[200] $1eh$ states).

\subsection{Heavy oblate supported clusters}

The low-energy spectrum in heavy clusters is expected to be
involved. In general, this is indeed the case. However,
as is seen from Fig. 4 (second and third columns), the picture
essentially depends on the reaction. Some responses
reveal pronounced structures. We will analyse
these structures by using
Fig. 5 as a guide. Though Fig. 5 shows the level scheme for the
small cluster, it hints the principle trends.

The first line of Fig. 4 displays the quadrupole
photoabsorption. We see  bunching of the strength
at 0.5-1.0 eV. The bunching involves two kinds of  quadrupole
transitions: i) between neighbouring subshells (like
E20([220]-[200]) in Fig. 5) and ii) inside the valence
subshell (like E22([220]-[202]) in Fig. 5).
The second line of Fig. 4 exposes photoabsorption for the
scissors mode. The scissors peak rises and
blue-shifts with the deformation. Unfortunately, both LEQM and
scissors modes cannot be detected in photoabsorption
and their plots are given here mainly for better understanding of
the subsequent STIRAP results.

The RF E10 plots demonstrate the overwhelming dominance of
$\lambda\mu =21$ mode over 20. This can be explained by
statistical arguments. First, the mode $\lambda\mu =21$
covers two projections $\mu =\pm 1$ instead of one in
$\lambda\mu =20$. Second, the mode $\lambda\mu =21$
usually has much more transition single-particle matrix
elements than 20.

The RF E10 response is presented by  two distinct peaks.
The first one originates from the deformation splitting
(like E21([211]-[202]) in Fig. 5) and exhibits a blue-shift with
increasing $|\delta_2|$. This peak is not seen
in E2 photoabsorption. At the same time, it displays
the correlation with the scissors mode. The second peak at
$\sim 0.6-0.7$ eV is mainly determined by E21 transitions
between neighbouring subshells (like E21([211]-[200]) in Fig. 5)
and its energy does not depend on the deformation.

In RF E11 case, the mode 22 comes to play. The response
is entangled at $\delta_2=-0.3$ but exhibits
a pronounced low-energy ($\sim 0.1$ eV) $\lambda\mu =$22 structure at
$\delta_2=-0.5$. For the first glance, this structure looks puzzling
since it cannot be explained by any option discussed above.
Indeed, its energy does not visibly depend on
the deformation splitting (like E22([220]-[202]
in Fig. 5 or 22 bump in Fig. 4, first line). It also has too low
energy to be explained by transitions between the neighbouring
subshells. Our analysis has revealed very specific origin of this
structure. It is explained
in Fig. 6 as a result of bunching $[Nn_z\Lambda ]$ levels
with the same $n_z$ (see E22 transitions $[523] \to [521]$
and  $[532] \to [530]$).  The bunching signifies that in
large systems with strong deformation the {\it asymptotic}
Nilsson-Clemenger quantum number $n_z$ (number of quants along
the symmetry axis $z$) becomes {\it exact}.

\subsection{STIRAP discussion}

Some essential points concerning detection of LEQM in STIRAP
should be commented.

The lifetime of the dipole plasmon is mainly determined
by the Landau fragmentation while the radiative decay
of the plasmon usually plays a minor role. The advantage
of the STIRAP is that it strongly enhances the population of
the desirable low-energy states in spite of a weak
natural radiative decay to them from  intermediate states.

To distinguish in STIRAP the electron
LEQM from possible excitations of other nature (contributions of
impurities, ionic modes, etc), one can use the feature of electronic
excitations to exhibit the inverse dependence of their energy
on cluster size.

It worth noting that STIRAP deals with both free and supported
systems \cite{Berg}.

\section{Conclusions}

The properties of the low-energy quadrupole modes
(LEQM) in deformed clusters were analysed within the
Kohn-Sham LDA RPA approach. The stimulated Raman adiabatic
passage (STIRAP) was considered as the most suitable
experimental method to detect LEQM. The population
of the LEQM in STIRAP was calculated for two special cases:
free light deformed clusters and supported heavy strongly-oblate
clusters. In the first case, LEQM spectrum is dilute and
can be resolved in STIRAP. The spectrum is easily associated
with particular electron-hole pairs. This finally allows to determine
the single-particle energies near the Fermi level.
Besides, close connection between the $\lambda\mu =$21
and scissors modes allows to observe the latter.

In heavy clusters, STIRAP population through the $\lambda\mu =10$
branch of the dipole plasmon is strictly dominated by two
$\lambda\mu =$21 peaks. The nature of the peaks is explained.
In the  population through the $\lambda\mu =11$ branch,
the soft $\lambda\mu =$22 mode devotes a special
attention. It originates from
the fact that in large systems with strong deformation the
{\it asymptotic} Nilsson-Clemenger quantum number $n_z$
becomes {\it exact}.

It worth noting, that the LEQM properties described above are
of a general character and so should take place not only in
sodium clusters considered here but also in other
clusters with quantum shells, for example in clusters of noble metals.
Since any other information on electron LEQM in deformed
clusters is absent, our study can be used as a first guide
in the field.

\begin{acknowledgments}
The work was supported by the Visitors Program of Max
Planck Institute for the
Physics of Complex Systems (Dresden, Germany). We thank
professors E. Duval and J.-M. Rost for valuable discussions.
\end{acknowledgments}

\newpage

{\large\bf Figure captions}:

\vspace{0.5cm}\indent
{\bf Figure 1}:
Scheme of population of the low-energy
quadrupole states $\lambda\mu =$20, 21 and 22 in RS, RF and
STIRAP reactions through $\lambda\mu =10$ (left)
and 11 (right)
branches of the dipole plasmon.

\vspace{0.5cm}\indent
{\bf Figure 2}:
Photoabsorption cross section for the dipole
plasmon in Na$_{15}^+$, Na$_{19}^+$ and  Na$_{27}^+$.
Parameters of quadrupole and hexadecapole deformations
are given in boxes. The experimental data
\protect\cite{SH} (triangles) are compared with our
results given as bars (for every RPA state)
and the strength function smoothed by the Lorentz weight
(with averaging parameter 0.25 eV).
Contributions of $\lambda\mu =10$ and 11
dipole branches (the latter is twice stronger) are
given by dashed curves. The bars are given in $eV \AA^2$.

\vspace{0.5cm}\indent
{\bf Figure 3}:
The dipole plasmon in Na$_{118}$
at oblate deformations $\delta_2=$-0.1 (first line),
$\delta_2=$-0.3 (second line), and  $\delta_2=$-0.5
(third line). Contributions of $\lambda\mu =10$ and 11
dipole branches (the latter is twice stronger) are
given by dashed curves. The photoabsorption is
smoothed by the Lorentz weight
with $\Delta =0.25$ eV.

\vspace{0.5cm}\indent
{\bf Figure 4}:
Low-energy electron modes in Na$^+_{15}$ (left
column ) and Na$_{118}$ at $\delta_2=$-0.3 (middle column) and
$\delta_2=$-0.5 (right column). The plots exhibit LEQM
photoabsorption  (first line), scissors M1 photoabsorption
(second line), STIRAP population of LEQM through $\lambda\mu =10$
(third line) and 11 (forth line) dipole branches. LEQM
are depicted by solid ($\lambda\mu =20$), dashed
($\lambda\mu =21$), and dotted ($\lambda\mu =22$)
curves. All the responses are smoothed by the Lorentz weight
with $\Delta =0.1$ eV.

\vspace{0.5cm}\indent
{\bf Figure 5}:
The electron level scheme for Na$^+_{15}$
in the spherical limit (left) and at the equilibrium
deformation (right). The Fermi level is [211].
Arrows depict the possible low-energy quadrupole
hole-electron $E2\mu$ transitions.

\vspace{0.5cm}\indent
{\bf Figure 6}:
Deformation splitting of the electron
subshell $2f$ in Na$_{118}$. The oblate deformations
are listed below the plot. The levels are marked by
Nilsson-Clemenger quantum numbers $[Nn_z \Lambda ]$.
The plot demonstrates
the origin of the soft E22 transitions.

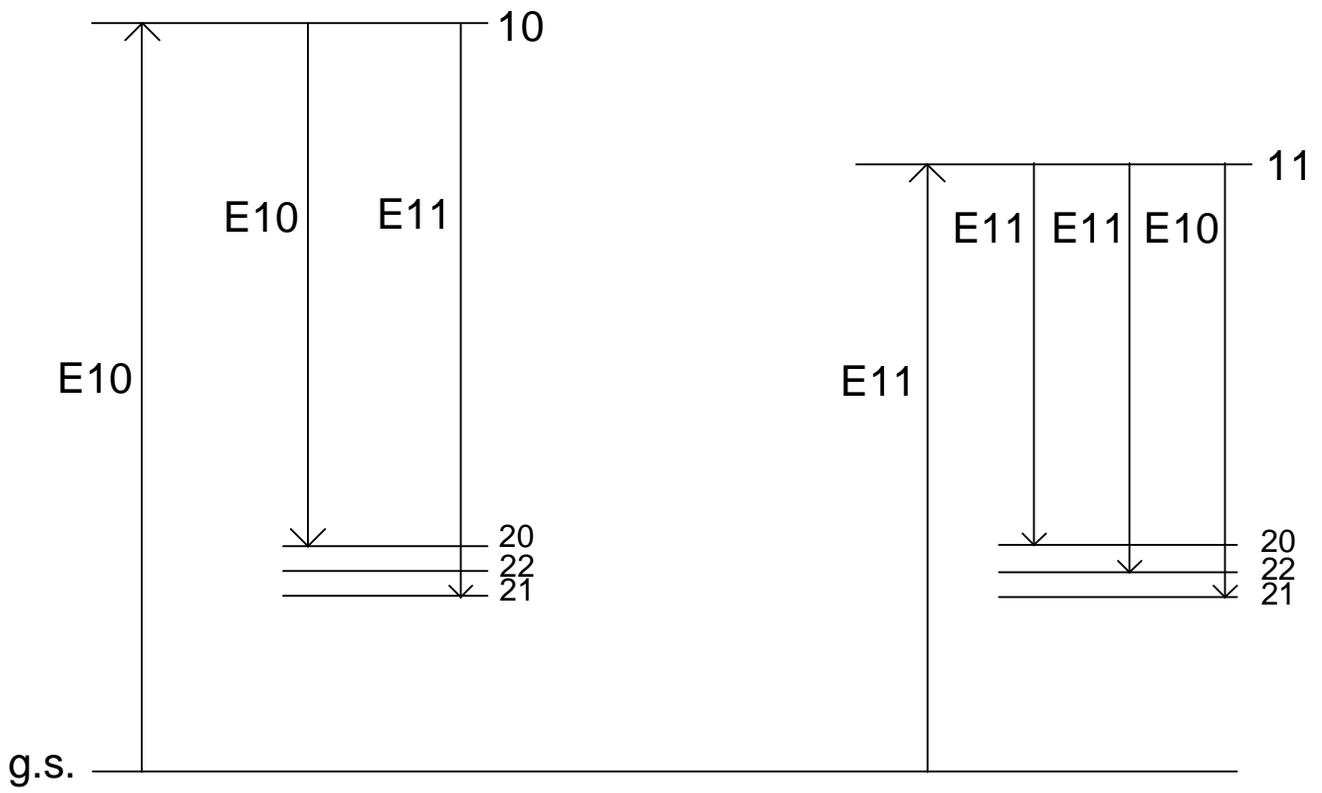
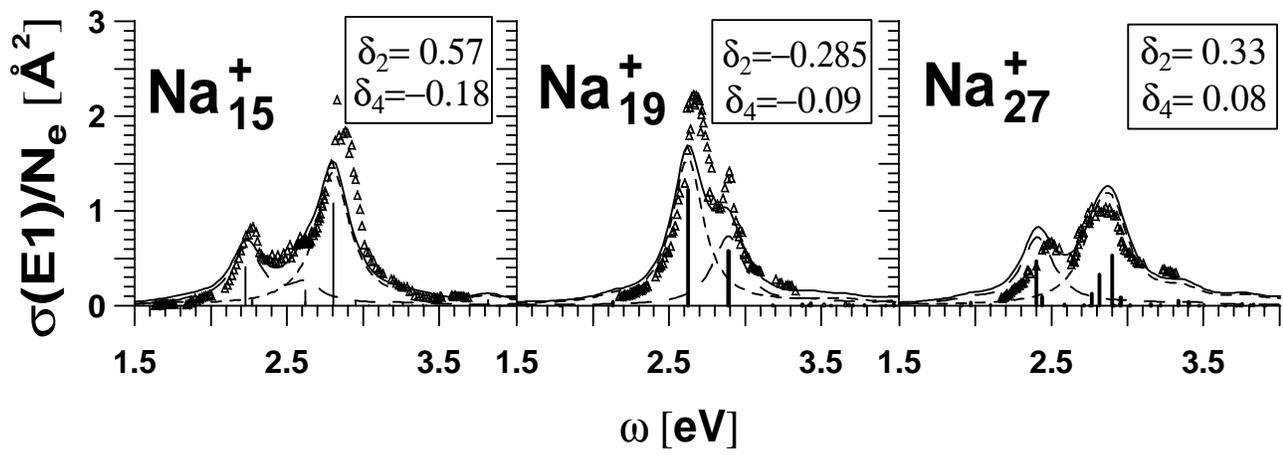
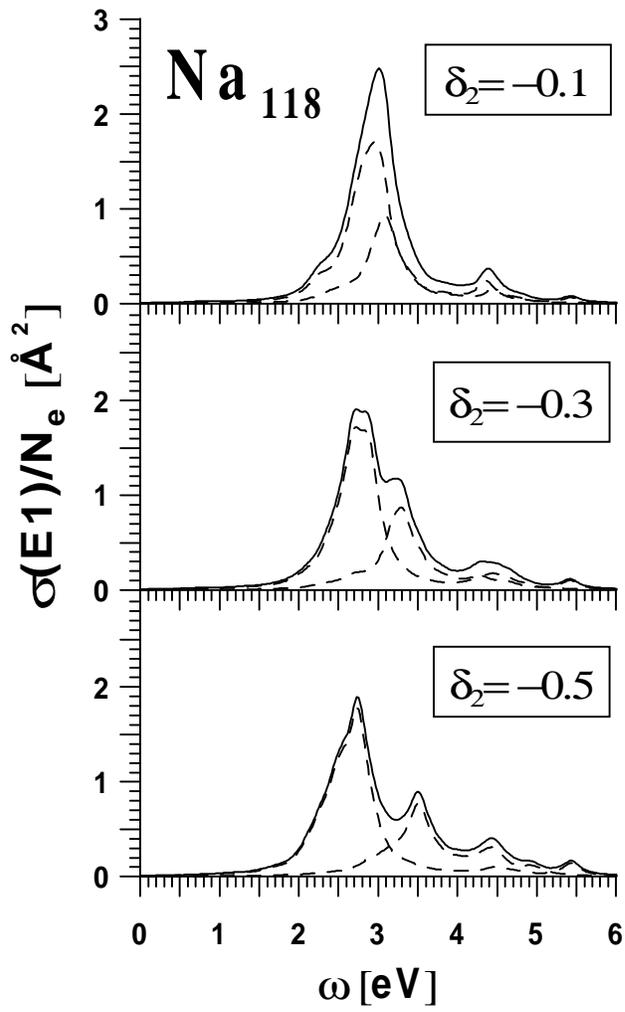
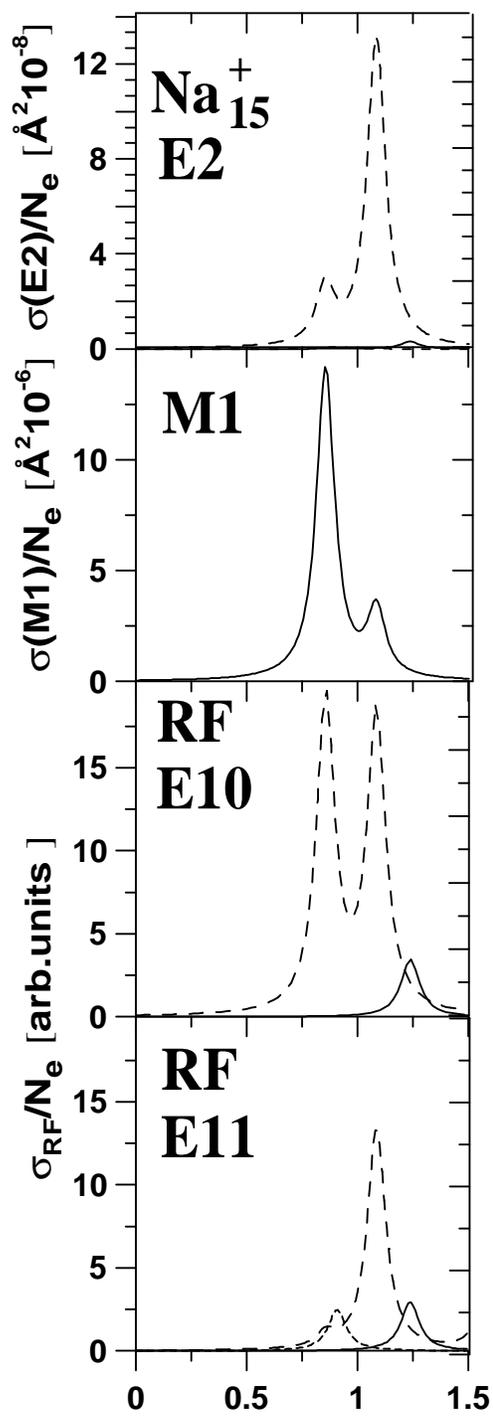
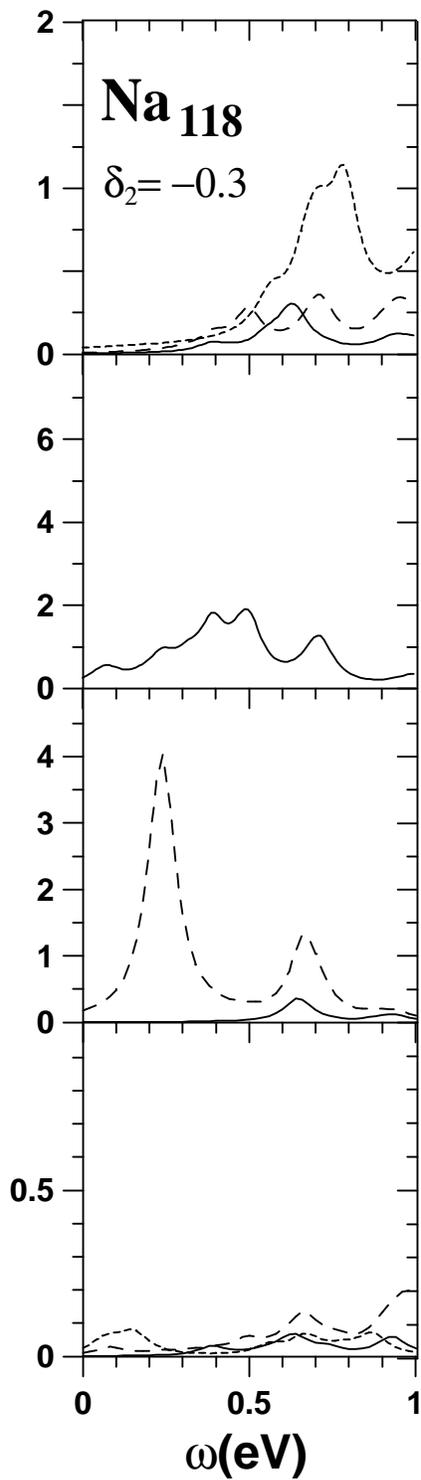
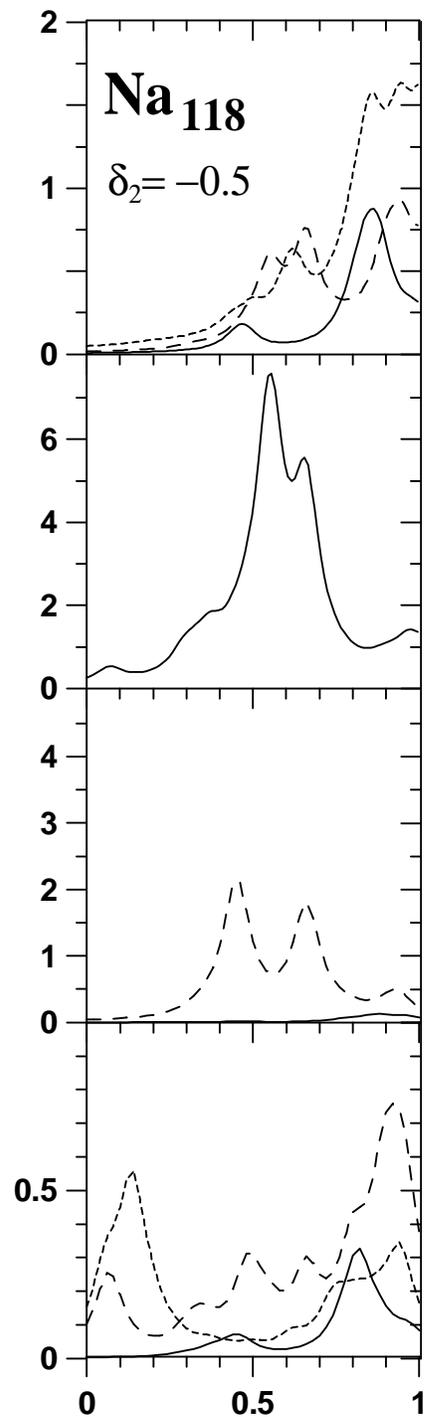
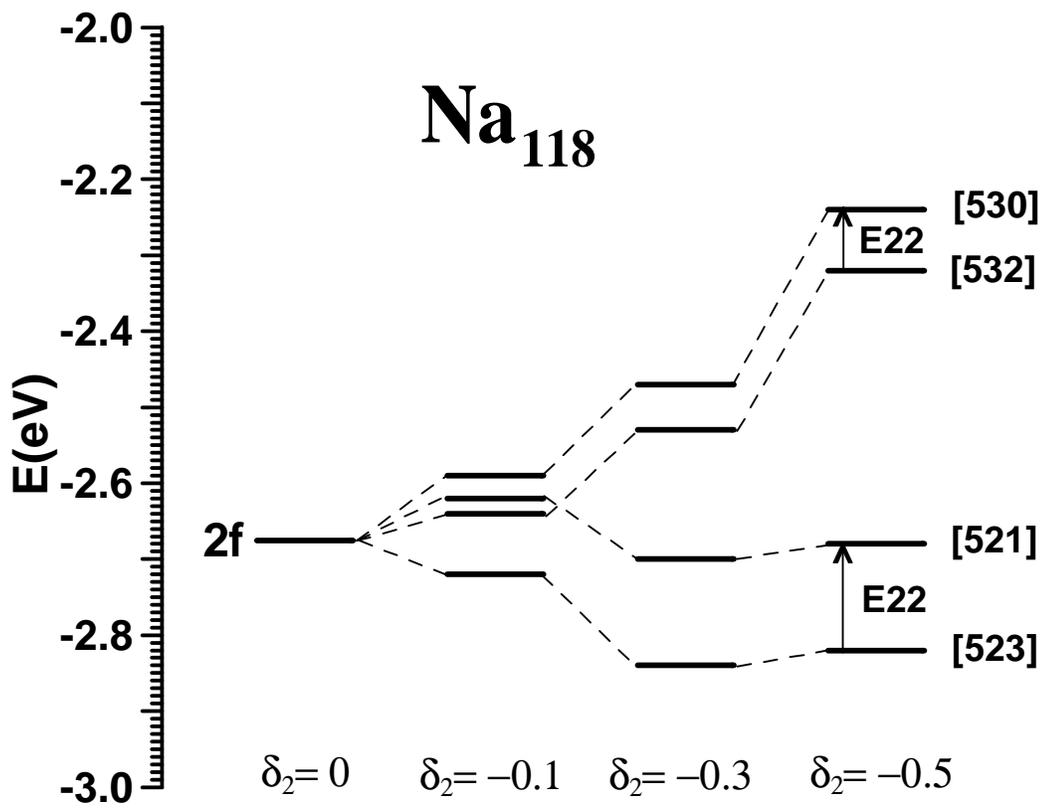
\end{document}